\documentclass[aps,pra,draft,showpacs,twocolumn]{revtex4}
\usepackage{graphicx}
\begin{document}

\title{The relation between Hardy's non-locality and violation of Bell inequality}

\author{Yang Xiang}
\email{njuxy@sohu.com}
\affiliation{School of Physics and Electronics, Henan University, Kaifeng 475001, Henan province, China }
\date{\today}
\begin{abstract}
We give a analytic quantitative relation between Hardy's non-locality and Bell operator.
We find that Hardy's non-locality is a sufficient condition for violation of Bell inequality, the upper bound
of Hardy's non-locality allowed by information causality just correspond to Tsirelson bound of Bell inequality, and
the upper bound of Hardy's non-locality allowed by the principle of no-signaling just correspond to the algebraic maximum of Bell operator.
Then we study the Cabello's argument of Hardy's non-locality (a generalization of Hardy's argument) and find a similar relation
between it and violation of Bell inequality. Finally, we give a simple derivation of the bound of Hardy's non-locality under the constraint of information
causality with the aid of above derived relation between Hardy's non-locality and Bell operator.

\end{abstract}
\keywords{Hardy's non-locality; Bell inequality; Information causality}

\pacs{03.65.Ud, 03.65.Ta}
\maketitle

%\keywords{}


\section{Introduction}
Quantum non-locality is always a fundamental problem in physics research. Violation of Bell inequality
claims that quantum non-locality cannot be reproduced with the hidden variable local model \cite{bell,chsh}.
In $1992$ Hardy proposed his theorem (Hardy's non-locality) which is a manifestation of quantum non-locality without using inequality \cite{hardy1,hardy2}.
However the amount of quantum non-locality is limited by Tsirelson bound \cite{tsir}, and the whole boundary of
quantum non-locality of binary-input and binary-output model has also been studied in \cite{nav1,nav2}. Tsirelson bound
(include the boundary of quantum non-locality) doesn't
originate solely from the principle of no-signaling, one class of no-signaling theory, initiated by Popescu and Rohrlich (PR correlation) \cite{pop},
allow the Bell operator take its algebraic maximum which greatly exceed Tsirelson bound. So identifying the physical
principles underlying the limits of quantum non-locality is now a intriguing problem in foundational research of QM. In 2005,
W. van Dam \cite{dam} has showed that the PR correlation makes communication complexity trivial (communication complexity is not trivial in QM),
but which seems highly implausible in nature. Recently, Paw{\l}owski \emph{et al.} \cite{pawlowski2} introduced a new physical principle, which is named
information causality, this new principle state that communication of $m$ classical bits causes information gain of at most $m$ bits
, and they prove that this principle can distinguish physical theories from other no-signaling theories that are endowed with stronger correlations
than quantum physics's. A new progress has been made in \cite{bar,acin} recently, the authors have proved that the quantum non-locality originate
from the combination of the principle of no-signaling and local quantum measurement assumption.

In order to investigate quantum correlation, Barret \emph{et al.} \cite{barrett}
placed them in a general correlation theory---no-signaling theory. They found that no-signaling correlations form a polytope in
probabilities space, and this no-signaling polytope contains the quantum correlations as a subset. Their work provide an mathematical framework for
research of quantum non-locality. Recently, within this mathematical framework of no-signaling polytope, Ahanj \emph{et al.} \cite{ahanj} give a bound
on Hardy's non-locality under the constraint of information causality. In the present paper we will
study the relation between Hardy's non-locality and violation of Bell inequality with the help of no-signaling polytope.
We find that the Hardy's non-locality is a sufficient condition for violation of Bell inequality, the upper bound
of Hardy's non-locality limited by information causality just correspond to Tsirelson bound of Bell inequality, and
the upper bound of Hardy's non-locality allowed by the principle of no-signaling just correspond to the algebraic maximum of Bell operator.
Then we study the Cabello's argument of Hardy's non-locality and find a similar relation
between it and violation of Bell inequality. Finally, we give a simple derivation of the bound of Hardy's non-locality under the constraint of information
causality.

\section{No-signaling polytope and Hardy's non-locality}
In this section we give a brief introduction of no-signaling polytope and Hardy's non-locality.

{\it Non-signaling polytope}~Let us still consider the case in which Alice and Bob are each choosing from two input, each of them
has two possible outputs. We denote $X(Y)\in\{0,1\}$ and $a(b)\in\{0,1\}$ as Alice's(Bob's) observable and
outcome respectively. The joint probabilities $p_{ab|XY}$ form a table with $2^{4}$ entries, although these are
not all independent due to the constraints of normalization and the principle of no-signaling.
These constraints lead the entire joint probabilities to a convex subset in the form of a polytope in $2^{4}$-dimensional probabilities
vector space, one call this polytope as no-signaling polytope \cite{barrett}. No-signaling polytope is eight dimensional, have $24$
vertices, $16$ of which are local vertices and $8$ of which are nonlocal vertices (they all correspond to PR correlations).

The local vertices can be expressed as
\begin{equation}
p_{ab|XY}^{\alpha \beta \gamma \delta} =
\left\{\begin{array}{ccccc}
                   1, & {\rm if} & a &=& {\alpha}X \oplus {\beta},\\
                     &           & b &=& {\gamma}Y \oplus {\delta};\\
                   0, & {\rm else} & & &
                   \end{array}
             \right.
\label{local vertices}
\end{equation}
where $\alpha, \beta, \gamma, \delta \in \{ 0, 1\}$ and $\oplus$ denotes addition modulo $2.$\\

And the eight nonlocal vertices are:
\begin{equation}
p_{ab|XY}^{\alpha \beta \gamma } =
\left\{\begin{array}{ccccc}
                   \frac{1}{2}, & {\rm if} & a \oplus b &=& XY \oplus {\alpha}X \oplus {\beta}Y \oplus \gamma,\\
                   0, & {\rm else} & & &
                   \end{array}
             \right.
\label{nonlocal vertices}
\end{equation}
where $\alpha, \beta, \gamma \in \{ 0, 1\}$.

For the case of two inputs and two outputs, there are eight nontrivial facets of the local correlations and they
correspond to eight CHSH inequalities as well as eight nonlocal vertices respectively. Let define $\langle ij\rangle$
as:
\begin{eqnarray}
\langle ij\rangle=\sum^{1}_{a,b=0}{(-1)^{a+b}p_{ab|X=i,Y=j}}.
\label{average}
\end{eqnarray}
Then these eight CHSH inequalities can be express as the following inequalities:
\begin{eqnarray}
B_{\alpha,\beta,\gamma}\equiv&&(-1)^{\gamma}\langle 00\rangle+(-1)^{\beta+\gamma}\langle 01\rangle
+(-1)^{\alpha+\gamma}\langle 10\rangle\nonumber\\
&&+(-1)^{\alpha+\beta+\gamma+1}\langle 11\rangle\leq 2,
\label{general bell ineq}
\end{eqnarray}
where $\alpha,\beta,\gamma\in\{0,1\}$. We can find that the algebraic maximum of Bell operator is $B_{\alpha,\beta,\gamma}=4$,
each choice of $\alpha,\beta,\gamma$ corresponds to one nonlocal vertex of Eq. (\ref{nonlocal vertices}), thus there
is a one-to-one correspondence between the nonlocal vertices of no-signaling polytope and the CHSH inequalities. It is easy to check
that each nonlocal vertex return a value for the corresponding Bell operator of $B_{\alpha,\beta,\gamma}=4$.

{\it Hardy's non-locality}~Consider Alice and Bob share two spin-$1/2$ particles, denote $\{A,A'\}$ as the measurement set of Alice and
$\{B,B'\}$ as the measurement set of Bob, all outcomes of measurement only take values of $\pm1$. Now consider
the following joint probabilities:
\begin{eqnarray}
P(A = +1, B = +1) &=& q_{1}\\
P(A^{\prime} = -1, B = -1) &=& 0\\
P(A = -1, B^{\prime} = -1) &=& 0\\
P(A^{\prime} = -1, B^{\prime} = -1) &=& q_{2}
\end{eqnarray}
, when $q_{1}=0$ and $q_{2}>0$ above four equations represent Hardy's argument of Hardy's non-locality.
The general case of $0\leq q_{1}<q_{2}$ corresponding to the Cabello's argument which is a
generalization of Hardy's argument.
It can be proved that there is no
local realistic theory that can reproduce the predictions of Eqs. $(5)$-$(8)$ \cite{ahanj}. To show this, let us
consider that there are some local realistic states for which $A'=-1$ and $B'=-1$, this
correspond to the validity of Eq. $(8)$. For these local realistic states Eqs. $(6)$ and $(7)$ tell
that the outcomes of $A$ and $B$ must be equal to $+1$, so according to local realistic theory $P(A = +1, B = +1)$
should be at least equal to $q_{2}$ and this contradict $q_{1}<q_{2}$. However quantum entangle state can reproduce
Hardy's non-locality with suitable measurement setting, an example of the most general nonmaximally entangle state
which can produce Hardy's non-locality can be seen in \cite{gar}. So Hardy's non-locality (theorem) manifest the
contradiction without inequalities  between local realism theory and quantum mechanics.

In \cite{ahanj}, the authors have given the expressions of Hardy's non-locality (Hardy's argument and Cabello's argument)
in terms of vertices of no-signalling polytope.
Using the correspondence as that in \cite{ahanj}: $(X=0)\leftrightarrow
A,(X=1)\leftrightarrow A^{\prime},$ $(Y=0)\leftrightarrow
B,(Y=1)\leftrightarrow B^{\prime}$ and $a,b=0(1)\leftrightarrow
+1(-1)$, one can find that there are only five of the $16$ local vertices and one of the $8$ nonlocal
vertices satisfy Hardy's argument Eq. $(5)$-$(8)$ ($q_{1}=0$), these vertices are $p_{ab|XY}^{0001}$, $p_{ab|XY}^{0011}$, $p_{ab|XY}^{0100}$,
$p_{ab|XY}^{1100}$, $p_{ab|XY}^{1111}$ and $p_{ab|XY}^{001}$. So Hardy's argument of Eq. $(5)$-$(8)$ can be written
as a linear superposition of the above $6$ vertices \cite{ahanj}:
\begin{eqnarray}
p_{ab|XY}^{{\cal H}} &=& c_1 p_{ab|XY}^{0001} + c_2
p_{ab|XY}^{0011} + c_3 p_{ab|XY}^{0100} \nonumber \\
&&+ c_4 p_{ab|XY}^{1100}+ c_5 p_{ab|XY}^{1111} + c_6
p_{ab|XY}^{001}
\label{hardy box}
\end{eqnarray}
where $\sum_{i = 1}^{6} {c}_i = 1$.

It's easy to check from Eq. (\ref{hardy box})
that the success probability $q_{2}$ for Hardy's argument is given by $q_{2}=p_{11|11}^{{\cal H}}=\frac{c_{6}}{2}$.
We can find that under the no-signaling constraint, the maximum of $q_{2}$ is $1/2$ which is achieved when $c_{6}=1$ and
else $c_{i}$'s$=0$. This result has also been derived in \cite{chou,cere}. Recently Ahanj \emph{et al.} \cite{ahanj} gave
$q_{2 max}=\frac{\sqrt{2}-1}{2}$ under the constraint of information causality \cite{pawlowski2}.

Cabello's argument
can also be expressed as a linear superposition of the vertices which satisfy Eqs. $(5)$-$(8)$ in the case of
$0\leq q_{1}<q_{2}$ \cite{ahanj}:
\begin{eqnarray}
p_{ab|XY}^{{\cal C}} &=& p_{ab|XY}^{{\cal H}} + c_7
p_{ab|XY}^{0000} + c_8 p_{ab|XY}^{0010} + c_9 p_{ab|XY}^{1000}
\nonumber \\
&&+ c_{10} p_{ab|XY}^{1010} + c_{11} p_{ab|XY}^{110},
\label{cabello box}
\end{eqnarray}
where the Hardy's argument $p_{ab|XY}^{{\cal H}}$ is given in Eq. (\ref{hardy box}), and the
coefficients $c_{i}$'s satisfy the new normalization condition $\sum_{i = 1}^{11} {c}_i = 1$.
It is easy to calculate the success probability for Cabello's argument by using Eq. (\ref{cabello box})
\begin{eqnarray}
w&\equiv& q_{2}-q_{1}\nonumber\\
&=&p_{11|11}^{{\cal C}}-p_{00|00}^{{\cal C}}\nonumber\\
&=&\frac{c_{6}-c_{11}}{2}-c_{7}-c_{8}-c_{9}
\label{w}
\end{eqnarray}
Recently Ahanj \emph{et al.} \cite{ahanj} gave
$w_{max}=\frac{\sqrt{2}-1}{2}$ under the constraint of information causality \cite{pawlowski2}.

\section{Hardy's non-locality and Bell inequality}

Now we study the relation between the Hardy's non-locality and violation of Bell inequality.
We first discuss the relation between $q_{2}$ and Bell operator in the case of $q_{1}=0$ (Hardy's argument).
We notice that the expression of
Hardy's argument (Eq. (\ref{hardy box})) has only one non-local vertex $p_{ab|XY}^{001}$, so it's nature to adopt
the Bell inequality corresponding to this non-local vertex:
\begin{eqnarray}
B_{0,0,1}\equiv -\langle00\rangle-\langle01\rangle-\langle10\rangle+\langle11\rangle\leq 2.
\label{bell ineq}
\end{eqnarray}

By using Eq. (\ref{average}) and Eq. (\ref{hardy box}), we can get
\begin{eqnarray}
\langle00\rangle=-c_{1}-c_{2}-c_{3}-c_{4}+c_{5}-c_{6}\nonumber\\
\langle01\rangle=-c_{1}+c_{2}-c_{3}-c_{4}-c_{5}-c_{6}\nonumber\\
\langle10\rangle=-c_{1}-c_{2}-c_{3}+c_{4}-c_{5}-c_{6}\nonumber\\
\langle11\rangle=-c_{1}+c_{2}-c_{3}+c_{4}+c_{5}+c_{6}.\nonumber\\
\label{00}
\end{eqnarray}
Then we obtain the value of Bell operator $B_{0,0,1}$
\begin{eqnarray}
B_{0,0,1}&=&2c_{1}+2c_{2}+2c_{3}+2c_{4}+2c_{5}+4c_{6}\nonumber\\
&=&2c_{6}+2\nonumber\\
&=&4q_{2}+2
\label{bell operator1}
\end{eqnarray}
In above calculation we used the normalization condition of $\sum_{i = 1}^{6} {c}_i = 1$ and the relation of $q_{2}=\frac{c_{6}}{2}$.
In 2000, Cereceda derived this result in \cite{cere}.

From Eq. (\ref{bell operator1}) we find that, if the success probability $q_{2}>0$ the
violation of Bell inequality can be achieved, and it can be also said that the Hardy's non-locality is a sufficient
condition for violation of Bell inequality. We also find that when $q_{2}=\frac{\sqrt{2}-1}{2}$ the Bell operator $B_{0,0,1}$ reach
Tsirelson bound $2\sqrt{2}$ and this value ($q_{2}=\frac{\sqrt{2}-1}{2}$) just equal to the upper bound of Hardy's non-locality under
the constraint of information causality \cite{ahanj}. But it must be noticed that in quantum mechanics the maximum probability of
success of the Hardy's non-locality is $q_{2}=0.09$ \cite{kunkri}, the reason is that there are also non-quantum correlations which under
Tsirelson bound $2\sqrt{2}$. Under the no-signaling constraint the maximum of $q_{2}$ is $1/2$, substitute it
in Eq. (\ref{bell operator1}) we just get the algebraic maximum of Bell operator: $B_{0,0,1}=4$.

Then we discuss the relation between $w$ and Bell operator in the case of $0<q_{1}<q_{2}$ (Cabello's argument).

By using Eq. (\ref{average}) and Eq. (\ref{cabello box}), we can get
\begin{eqnarray}
\langle00\rangle&=&-c_{1}-c_{2}-c_{3}-c_{4}+c_{5}-c_{6}\nonumber\\
&&+c_{7}+c_{8}+c_{9}+c_{10}+c_{11}\nonumber\\
\langle01\rangle&=&-c_{1}+c_{2}-c_{3}-c_{4}-c_{5}-c_{6}\nonumber\\
&&+c_{7}-c_{8}+c_{9}-c_{10}-c_{11}\nonumber\\
\langle10\rangle&=&-c_{1}-c_{2}-c_{3}+c_{4}-c_{5}-c_{6}\nonumber\\
&&+c_{7}+c_{8}-c_{9}-c_{10}-c_{11}\nonumber\\
\langle11\rangle&=&-c_{1}+c_{2}-c_{3}+c_{4}+c_{5}+c_{6}\nonumber\\
&&+c_{7}-c_{8}-c_{9}+c_{10}-c_{11}
\label{11}
\end{eqnarray}

Here we still adopt the Bell inequality
of Eq. (\ref{bell ineq}) which corresponding to non-local vertex $p_{ab|XY}^{001}$, we obtain the value of Bell
operator $B_{0,0,1}$ for Cabello's argument of Eq. (\ref{cabello box})
\begin{eqnarray}
B_{0,0,1}&=&-\langle00\rangle-\langle01\rangle-\langle10\rangle+\langle11\rangle\nonumber\\
&=&2+2c_{6}-2c_{11}-4(c_{7}+c_{8}+c_{9})\nonumber\\
&=&2+4w
\label{bell operator2}
\end{eqnarray}
In above calculation we used the normalization condition of $\sum_{i = 1}^{11} {c}_i = 1$ and the relation of Eq. (\ref{w}).

From Eq. (\ref{bell operator2}) we find a similar relation between $w$ and violation of Bell inequality as that of $q_{2}$'s.
If the success probability for Cabello's argument $w=q_{2}-q_{1}>0$ the violation of Bell inequality can be achieved.
One also can find that when $w=\frac{\sqrt{2}-1}{2}$ the Bell operator $B_{0,0,1}$ reach
Tsirelson bound $2\sqrt{2}$ and this value ($w=\frac{\sqrt{2}-1}{2}$) just equal to the upper bound of Cabello's non-locality under
the constraint of information causality \cite{ahanj}. Under the no-signaling constraint, the maximum of $w$ is $1/2$ when
$c_{6}=1$ and rest of the $c_{i}$'s$=0$, substitute it
in Eq. (\ref{bell operator2}) we just get the algebraic maximum of Bell operator of $B_{0,0,1}=4$.

\section{The bound of Hardy's non-locality in the limits of information
causality}
In this section, we give a simple derivation of the bound of Hardy's non-locality under the constraint of information
causality with the aid of above derived relation between Hardy's non-locality and Bell operator.

The general Bell inequality of Eq. (\ref{general bell ineq}) can also be written as
\begin{eqnarray}
B^{'}_{\alpha,\beta,\gamma}&=&\frac{1}{4}\sum_{X,Y=0}^{1}{p(a=b\oplus XY\oplus\alpha X\oplus\beta Y\oplus\gamma|XY)}\nonumber\\
&\leq&\frac{3}{4},
\label{new bell ineq}
\end{eqnarray}
where $\alpha,\beta,\gamma \in\{0,1\}$. The choice of $\alpha=0,\beta=0,\gamma=0$ corresponding to the standard CHSH inequality \cite{chsh}
and be widely used \cite{bar,acin}.
The relation between $B^{'}_{\alpha,\beta,\gamma}$ and $B_{\alpha,\beta,\gamma}$
is
\begin{eqnarray}
B^{'}_{\alpha,\beta,\gamma}&=&\frac{B_{\alpha,\beta,\gamma}+4}{8}\nonumber\\
&=&\frac{4q_{2}+6}{8}~~~~(Hardy's~~argument)\nonumber\\
&=&\frac{4w+6}{8}~~~~(Cabello's~~argument).
\label{bb}
\end{eqnarray}
Tsirelson bound of $B^{'}_{\alpha,\beta,\gamma}$ is $\frac{\sqrt{2}+2}{4}$
when $q_{2}=\frac{\sqrt{2}-1}{2}$ (or $w=\frac{\sqrt{2}-1}{2}$).

The general expression of information causality can written as
\begin{eqnarray}
A\equiv(2P_{1}-1)^{2}+(2P_{2}-1)^{2}\leq1
\label{information causality}
\end{eqnarray}
where
\begin{eqnarray}
P_{1}&=&\frac{1}{2}\big[p(a=b\oplus XY\oplus\alpha X\oplus\beta Y\oplus\gamma|00)\nonumber\\
&&+p(a=b\oplus XY\oplus\alpha X\oplus\beta Y\oplus\gamma|10)\big]\nonumber\\
P_{2}&=&\frac{1}{2}\big[p(a=b\oplus XY\oplus\alpha X\oplus\beta Y\oplus\gamma|01)\nonumber\\
&&+p(a=b\oplus XY\oplus\alpha X\oplus\beta Y\oplus\gamma|11)\big].\nonumber\\
\label{pp}
\end{eqnarray}
the expression of this principle in \cite{pawlowski2} correspond to the choice of $\alpha=0,\beta=0,\gamma=0$.

The $B^{'}_{\alpha,\beta,\gamma}$ can be written as $\frac{P_{1}+P_{2}}{2}$ therefore
which can expressed as a function of $A$:
\begin{eqnarray}
B^{'}_{\alpha,\beta,\gamma}&=&\frac{(\sqrt{A}\sin{\theta}+\sqrt{A}\cos{\theta}+2)}{4}\nonumber\\
&\leq&\frac{(\sqrt{2A}+2)}{4}.
\label{condition}
\end{eqnarray}
We can find under the constraint of information causality $A\leq1$ the upper bound of $B^{'}_{\alpha,\beta,\gamma}$
is $\frac{\sqrt{2}+2}{4}$, therefore from the Eq. (\ref{bb}) we find the upper bound of $q_{2}$ and $w$ both
are $\frac{\sqrt{2}-1}{2}$ under the constraint of information causality. This upper bound of $q_{2}$ and $w$
can be achieved in the limits of information causality. For example we can take $P_{1}=P_{2}=\frac{\sqrt{2}+2}{4}$,
substitute them in Eq. (\ref{information causality}) we get $A=1$, so information causality has been observed;
at the same time we take them in Eq. (\ref{new bell ineq}) and get $B^{'}_{\alpha,\beta,\gamma}=\frac{1}{2}(P_{1}+P_{2})=\frac{\sqrt{2}+2}{4}$,
so from Eq. (\ref{bb}) we can find the upper bound of $q_{2}$ and $w$ ($\frac{\sqrt{2}-1}{2}$) has been reached.
Now we complete the proof of that the bound of Hardy's non-locality under the constraint of information causality
is $\frac{\sqrt{2}-1}{2}$, which is same as the result in \cite{ahanj}.

\section{Conclusion}

In this work, we give the quantitative relations between Hardy's/Cabello's argument of non-locality and violation of Bell inequality.
We obtain the analytic expressions of the relations between the success probabilities
of the Hardy's/Cabello's argument and the value of Bell operator, and then we find that if and
only if the success probabilities of the Hardy's/Cabello's
argument are greater than zero the violations of Bell inequality can be achieved. The bound values of the success probabilities of the Hardy's/Cabello's
argument under the constraint of information causality both correspond to Tsirelson bound of Bell operator, and the bound values of these two
success probabilities under the no-signaling constraint both correspond to the algebraic maximum of Bell operator.
Finally, we give a simple derivation of the bound of Hardy's non-locality under the constraint of information
causality with the aid of above derived relation between Hardy's non-locality and Bell operator, this bound is the main result of reference \cite{ahanj}.

\vskip 0.5 cm

\section*{Acknowledgments}

The author thank Wei Ren and Shi-Jie Xiong for their help and encouragement.
This work was supported by National Foundation of Natural Science in
China Grant Nos. 10947142 and 11005031.

%\end{acknowledgments}

%\bigskip

\end{document}